\documentclass{article}
\usepackage{graphicx}
\graphicspath{%
    {converted_graphics/}
    {C:/Dropbox/1-kgl-top/Papers/2015/SurfaceDiffusion/}
}
\begin{document}

\begin{center}
{\LARGE Incorporating Surface Diffusion into a Cellular Automata Model}\vskip%
6pt

{\LARGE of Ice Growth from Water Vapor}\vskip12pt

{\Large Kenneth G. Libbrecht}\vskip5pt

{\large Department of Physics, California Institute of Technology}\vskip2pt

{\large Pasadena, California 91125}\vskip-1pt

\vskip18pt

\hrule\vskip1pt \hrule\vskip14pt
\end{center}

\textbf{Abstract.} We describe a numerical model of faceted crystal growth
using a cellular automata method that incorporates admolecule diffusion on
faceted surfaces in addition to bulk diffusion in the medium surrounding the
crystal. The model was developed for investigating the diffusion-limited
growth of ice crystals in air from water vapor, where the combination of
bulk diffusion and strongly anisotropic molecular attachment kinetics yields
complex faceted structures. We restricted the present model to cylindrically
symmetric crystal growth with relatively simple growth morphologies, as this
was sufficient for making quantitative comparisons between theoretical
models and ice growth experiments. Overall this numerical model reproduces
ice growth behavior with reasonable fidelity over a wide range of
conditions, albeit with some limitations. The model could easily be adapted
for other material systems, and the cellular automata technique appears well
suited for investigating crystal growth dynamics when strongly anisotropic
surface attachment kinetics cause faceted growth morphologies.

\section{Introduction}

The formation of crystalline structures during solidification yields a
remarkable variety of morphological behaviors, resulting from the often
subtle interplay of non-equilibrium physical processes over a range of
length scales. In many cases, seemingly small changes in surface molecular
structure and dynamics at the nanoscale can produce large morphological
changes at all scales. Some examples include free dendritic growth from the
solidification of melts, where small anisotropies in the interfacial surface
energy govern the overall characteristics of the growth morphologies \cite%
{dendrites94, brener96}, whisker growth from the vapor phase initiated by
single screw dislocations and other effects \cite{whiskers07}, the formation
of porous aligned structures from directional freezing of composite
materials \cite{directional05}, and a range of other pattern formation
systems \cite{cross93, kassner96}. Since controlling crystalline structure
formation during solidification has application in many areas of materials
science, much effort has been directed toward better understanding the
underlying physical processes and their interactions.

We have been exploring the growth of ice crystals from water vapor in an
inert background gas as a case study of how complex faceted structures
emerge in diffusion-limited growth. Although this is a relatively simple
monomolecular physical system, ice crystals exhibit columnar and plate-like
growth behaviors that depend strongly on temperature, and much of the
phenomenology of their growth remains poorly understood \cite%
{libbrechtreview05, nelson01, pruppacher97}. Ice has also become something
of a standard test system for investigating numerical methods of faceted
crystal growth \cite{reiter05, gg09, garcke12, kelly13}. A better
understanding of ice crystal formation yields insights into the detailed
molecular structure and dynamics of the ice surface, which in turn
contributes to our understanding of many meteorological, biological, and
environmental processes involving ice \cite{clouds04, moldymice02, dash06}.

In our investigation of how surface energy and attachment kinetics affect
ice growth dynamics, we needed a quantitative numerical model that would
allow us to \textquotedblleft grow\textquotedblright\ model ice crystals for
comparison with experimental measurements of growth rates and morphologies.
Although proven numerical methods for modeling diffusion-limited growth have
been available for years, many of the existing methods are ill-suited for
modeling ice growth behavior. For example, phase-field \cite{phasefield06,
phasefield08} and front-tracking \cite{schmidt96} methods have demonstrated
the ability to accurately model diffusion-limited growth for the case of
fast attachment kinetics and a weakly anisotropic surface energy, which is
characteristic of most solidification from the melt. These systems typically
yield unfaceted dendritic structures, however, in contrast to strongly
faceted ice structures. Early models for the growth of faceted crystals \cite%
{yokoyama93, baker01} were generally too limited to allow quantitative
comparisons with ice growth data.

Modeling diffusion-limited growth in systems with strong surface
anisotropies has proven difficult, and only recently have researchers
demonstrated robust techniques capable of generating structures that are
both faceted and dendritic. Reiter \cite{reiter05} described an especially
promising cellular automata simulator that solves the diffusion equation by
nearest neighbor relaxation, including a set of parameterized nearest
neighbor rules to define the boundary conditions at the crystal interface.
This model was further advanced by several researchers \cite{gg06, gg08,
gg09, libbrecht08, kglca13, kelly13}, and the method yields a deterministic
dendritic growth behavior in which faceting follows the symmetry of the
predefined numerical grid.

Barrett et al. \cite{garcke12} also developed a robust adaptive mesh
technique that generated faceted dendritic crystal growth patterns. In this
work the authors found that a strongly anisotropic surface energy was
required to produce faceted dendritic growth, while anisotropic attachment
kinetics alone were not sufficient to reproduce this behavior. We have
suggested that the ice case is more likely described by the opposite
characteristics -- a nearly isotropic surface energy together with strongly
anisotropic attachment kinetics, the latter dominating the growth behavior 
\cite{ecs12}. In fact, the roles played by these two physical effects are
not yet known with certainty.

The relative merits of different computational methods for modeling
diffusion-limited growth in the presence of strong surface anisotropies are
not presently well understood, as this is an area of current research.
Moreover, our knowledge of the surface physics governing the growth of
faceted materials is itself rather poor, including the relative
contributions of the anisotropies in surface energy and attachment kinetics
in different materials. In our experience, progress on both these research
fronts is linked: better modeling methods allow more accurate interpretation
of growth experiments, in turn fostering improved experiments that yield a
better understanding of the surface physics input into the models.

Below we describe a cellular automata method for modeling diffusion-limited
growth in the presence of strongly anisotropic molecular attachment
kinetics, focusing on ice growth from water vapor. The model is an extension
of that presented in \cite{kglca13}, now including admolecule diffusion on
faceted surfaces, which is quite important when vicinal surfaces are
present. While we are not yet able to reproduce all aspects of faceted
growth behavior, our model is robust, numerically well-behaved,
computationally straightforward, and quite flexible for exploring ice growth
behaviors.

The 2D cylindrically symmetric model has been especially useful for
investigating the simple growth morphologies often produced in experiments.
A basic hexagonal prism, for example, is modeled by a right cylinder,
replacing the six prism facets with a single cylindrical \textquotedblleft
facet\textquotedblright . This model is adequate for basic plate and needle
morphologies, as well as some more complex forms such as capped columns,
hollow columns, and double plates. Using a 2D model allows the rapid
generation of hundreds of model crystals for comparison with experimental
results, using different input assumptions. We have found that this model is
quite useful when examining the surface physical processes governing ice
growth rates, and it allows straightforward adaptation for use in other
investigations involving faceted diffusion-limited crystal growth.

\section{Incorporating Surface Diffusion}

The new model presented here is a direct extension of what we described in
detail in \cite{kglca13}, so we will not reproduce a derivation of the
physics underlying the central features of the cellular automata method. We
reiterate, however, that our overarching goal in this effort has been to
define a physically accurate crystal growth model for \textit{quantitative}
comparison with crystal growth experiments. Thus we strive to model not only
the correct faceted crystal morphologies, but correct crystal growth rates
as well. This is in contrast to several earlier cellular automata models of
ice growth \cite{reiter05, gg06, gg08, gg09}, in which particle diffusion in
the medium surrounding the crystal was treated correctly, but with somewhat 
\textit{ad hoc} surface boundary conditions.

To date, all the cellular automata models of ice crystal growth have
incorporated local surface boundary conditions. In these models, the
boundary conditions to the bulk diffusion equation at a given point on the
ice surface depend only on conditions at that point, and are independent of
processes occurring at other surface locations. Such local models do not
include admolecule diffusion on faceted surfaces, as surface transport is a
nonlocal process. As described in \cite{kglca13}, we previously assumed that
surface diffusion processes that act over length scales of order $x_{s},$
the surface diffusion length \cite{saito96}, could be incorporated into the
local attachment coefficient $\alpha ,$ as long as the model resolution was
restricted to length scales greater than $x_{s}.$ We next show that this
assumption was incorrect.

\subsection{A Microscopic Model}

Before examining the effects of surface diffusion in our macroscopic
cellular automata model, we first define an appropriate molecular model of
the process \cite{saito96}. Note that our model of surface diffusion is
independent of the bulk diffusion taking place in the medium surrounding the
crystal. In our physical picture of ice crystal growth from water vapor,
both diffusion processes are present: water molecules first diffuse through
the air to reach the ice crystal surface, and subsequently diffuse along the
crystal surface before becoming incorporated into the crystal lattice. (Heat
diffusion from latent heat deposition at the growing surface is neglected
for the reasons stated in \cite{libbrechtreview05}.) We consider only
admolecule diffusion on faceted surfaces. On rough surfaces, we retain the
assumption that admolecules are immediately incorporated into the lattice,
so the attachment coefficient is $\alpha \approx 1$ on nonfaceted surfaces.

\begin{figure}[htb] 
  \centering
  \includegraphics[width=5.67in,height=2.02in,keepaspectratio]{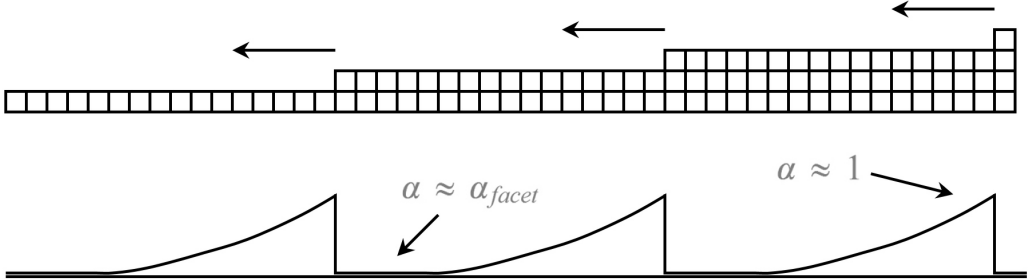}
  \caption{A basic molecular model of
surface diffusion, here depicted in 2D $(r,z)$ space. The boxes in the upper
diagram represents water molecules, collectively showing a cut-away side
view of a vicinal surface. The curve below indicates the attachment
coefficient $\protect\alpha $ along the surface. As described in the text, $%
\protect\alpha \approx \protect\alpha _{facet}$ far from kink sites, while $%
\protect\alpha \approx 1$ near kink sites. In this sketch the diffusion
distance $x_{s}$ is just a few molecules wide, while in reality $x_{s}$ may
be much larger. As the crystal grows, water molecules attach to the kink
sites, so the edge of each terrace advances to the left.}
  \label{vicinal}
\end{figure}

To examine surface diffusion in detail, consider a simple vicinal surface,
on which a series of molecular terrace steps are separated by a uniform
spacing $x_{step}$, as shown in Figure \ref{vicinal}. Let $x_{s}$ equal the
usual surface diffusion length \cite{saito96}, equal to the typical distance
admolecules diffuse during their residence time on the surface. We assume
for the present discussion that the Ehrlich--Schwoebel barrier reduces
admolecule diffusion over terrace steps to a negligible rate. If $%
x_{step}\ll x_{s}$, then essentially all admolecules will diffuse to kink
sites and be absorbed, contributing to crystal growth. The attachment
coefficient for such a vicinal surface is then $\alpha \approx 1;$ all
molecules that strike the surface become incorporated into the crystal
lattice.

If $x_{step}\gg x_{s},$ then only admolecules within a distance
approximately $x_{s}$ from each step will be absorbed, and $\alpha <1$ when
averaged over the surface. More specifically, we define an attachment
coefficient arising from surface diffusion as $\alpha _{SD}\approx \exp
(-\Delta x/x_{x}),$ where $\Delta x$ is the distance to an accessible kink
site \cite{saito96}, as shown in Figure \ref{vicinal}. For a faceted surface
near a terrace step, therefore, we must consider both $\alpha _{SD}$ and $%
\alpha _{facet},$ the latter being the intrinsic attachment coefficient for
a perfectly faceted surface.

Is is useful to frame this molecular process in the language of a cellular
automata model \cite{kglca13}, taking the cells in Figure \ref{vicinal} to
be molecule-size pixels in the numerical model. To include surface diffusion
in this model, we should: 1) include $\alpha _{SD}$ on facet sites that are
near kink sites, and 2) transfer the additional accumulated mass $dM$ --
that part arising from surface diffusion -- to the appropriate kink sites.
(See \cite{kglca13} for a definition of the accumulated mass in the cellular
automata model.)

In our numerical testing of this model, we have found that item (2) can be
ignored without substantially changing the growth rate or morphological
behavior. Making the assumption $\alpha _{SD}=\exp (-\Delta x/x_{s})$ from
surface diffusion at each point along the surface, with no mass transport to
kink sites, models the growth rate with reasonable fidelity. The reason for
this is that adding $dM$ just ahead of a moving kink site gives essentially
the same numerical result as transporting the same $dM$ to the kink site.
The key element of surface diffusion in our model is to increase $\alpha $,
and thus increase $dM,$ near kink sites.

The value of $x_{s}$ on faceted ice surfaces is not well known. In \cite%
{furukawa14}, the authors reported $x_{s}\approx 5$ $\mu $m for the
diffusion length on a basal facet at -8 C; however the same data were
reinterpreted in \cite{kglSD} to obtain a value $x_{s}\approx 10$ nm. This
is an area of active research, and there is a clear need for additional
measurements. Nevertheless, for vicinal surfaces tilted by angles as low as $%
\theta \approx a/x_{s}$ we can effectively assume $\alpha \approx 1$ on the
surface. Since $\alpha _{facet}$ can be quite small when $\theta =0,$ a
large $x_{s}$ would mean that that $\alpha \left( \theta \right) $ exhibits
an extremely sharp cusp at $\theta =0,$ pinpointing the difficulty inherent
in modeling faceted crystal growth.

\subsection{The Macroscopic Model}

We next apply this physical picture of surface diffusion to our 2D
cylindrically symmetric model in $(r,z)$ space \cite{kglca13}, which has
pixels that are typically $\Delta r=\Delta z\approx 0.15$ $\mu $m in size.
Our numerical algorithm for surface diffusion is essentially that described
in Figure \ref{vicinal} for the microscopic model. For each faceted boundary
pixel we calculate $\alpha _{SDleft}$ and $\alpha _{SDright}$ representing
the accumulated mass absorbed by kink sites on either side of a particular
facet site. If there is no kink site in one or both directions, then the
appropriate $\alpha _{SDx}$ values are zero. Specifically, for each facet
site we take%
\begin{eqnarray*}
\alpha _{SDleft} &=&A_{SD}\exp (-\Delta n_{left}/n_{SD}) \\
\alpha _{SDright} &=&A_{SD}\exp (-\Delta n_{right}/n_{SD})
\end{eqnarray*}%
where $\Delta n_{left}$ and $\Delta n_{right}$ are the distances to
accessible kink sites (if they exist) on either side of the facet site, in
pixels. For basal facets (in our 2D model), the $\Delta n_{x}$ are pixel
distances in the $r$ direction from the facet site in question; for prism
facets, the $\Delta n_{x}$ are pixel distances in the $z$ direction. The
constant $A_{SD}$ is typically set to unity, but we leave it as an
adjustable parameter in our model.

Our choice for $n_{SD}$ follows from the discussion of vicinal angles above,
from which we obtain $n_{SD}=(x_{s}/a)$ pixels. With this value for $n_{SD},$
we have $\alpha _{SD}\approx A_{SD}\approx 1$ whenever a vicinal surface is
tilted by an angle greater than $\theta \approx 1/n_{SD}\approx a/x_{s},$
the same angle as we found above. By this straightforward equal-angle
argument, we see that the effects of surface diffusion do not extend to $%
x_{s}/\Delta z$ pixels, as one might naively assume, but a factor of $%
(\Delta z/a)\approx 500$ times farther. This additional factor of $(\Delta
z/a)$ is a key result in this paper, as it shows the importance of including
surface diffusion in a cellular automata model of ice growth.

At this point we can see a possibly important limitation in our numerical
model. With a spatial resolution of $\Delta z,$ vicinal surfaces with $%
\theta <\Delta z/L$ will be indistinguishable from perfectly faceted
surfaces, where $L$ is the overall size of the facet. Such surfaces will
have no $\Delta z$ steps in the model, so must have $\alpha =\alpha
_{facet}\ll 1,$ In reality, however, the vicinal angle must go to $\theta
<a/L\ll \Delta z/L$ before $\alpha =\alpha _{facet}.$ If $x_{s}$ is large,
this means our model will have difficulty modeling nearly faceted surfaces,
which we will discuss in more detail below.

To complete our numerical algorithm of surface diffusion, we compute a total 
$\alpha _{SD}$ at each facet site using%
\begin{eqnarray*}
\alpha _{SD} &=&\alpha _{SDleft}+\alpha _{SDright}-\alpha _{SDleft}\alpha
_{SDright} \\
&=&C(\alpha _{SDleft},\alpha _{SDright})
\end{eqnarray*}%
(We note in passing that $\alpha _{SD}$ is calculated from the crystal
geometry alone, independent of the supersaturation field surrounding the
crystal.) We also compute $\alpha _{facet}$ for each facet site, which
generally depends on the surface supersaturation $\sigma _{surf}$, and this
is combined with $\alpha _{SD}$ to form the total attachment coefficient $%
\alpha _{tot}=C(\alpha _{facet},\alpha _{SD})$. This addition of $\alpha $
terms follows from the \textquotedblleft at least one\textquotedblright\
rule for combining independent probabilities, the general case being%
\[
p_{tot}=1-\prod_{i}(1-p_{i}) 
\]%
(M. Libbrecht, private communication). The reader can verify that this rule
gives the expected results for $\alpha _{tot}$ in various limits. For
example, $\alpha _{tot}\approx 1$ on the lower terrace near a step, $\alpha
_{tot}\approx \alpha _{facet}$ far from any step, and $\alpha _{tot}\approx
1 $ independent of $\alpha _{SD}$ if $\alpha _{facet}\approx 1$.

The result of including this surface diffusion algorithm can be seen by
considering the top sketch in Figure \ref{vicinal}, this time interpreting
the cells as pixels in the cellular automata model. Atop the highest terrace
on the facet surface, which we call the \textquotedblleft top
terrace\textquotedblright , we still have $\alpha =\alpha _{facet},$ because
there are no accessible kink sites (owing to our initial assumption of a
large Ehrlich--Schwoebel barrier). At all other sites, $\alpha _{tot}$ is
much increased, typically to $\alpha _{tot}\approx 1$ everywhere except on
the top terraces (because $n_{SD}=(\Delta z/a)x_{s}\approx 500\ $pixels is
typically larger than the crystals we are modeling, so $\exp (-\Delta
n_{left}/n_{SD})\approx 1)$. This is in contrast to our previous model
without surface diffusion \cite{kglca13}, where we had $\alpha =\alpha
_{facet}$ on all facet sites and $\alpha =1$ only on kink sites.

We can see how the addition of surface diffusion in our model promotes
faceting in two ways. First, lower terrace sites have larger $\alpha $
values and thus grow more quickly. Second, the larger $\alpha $ values
decrease the $\sigma $ field everywhere around the crystal, including on the
top terrace. Because $\alpha _{facet}$ typically depends strongly on $\sigma
_{surf},$ this lowers $\alpha _{facet}$ and thus reduces the nucleation of
new terraces on the top terrace. Thus the lower terraces fill in more
quickly, while nucleation of new terraces is reduced. These two factors both
lead to increased faceting.

\section{The 2D Cellular Automata Model}

It is beneficial at this point to describe the flow through our numerical
model in some detail. Here again, the reader can find the derivation of many
of the algorithms below described in \cite{kglca13}.

\subsection{Attachment Coefficients}

The intrinsic attachment coefficients for faceted prism and basal facets are
typically parameterized by $\alpha _{facet}=A\exp (-\sigma _{0}/\sigma
_{surf})$. where $\sigma _{surf}$ is the supersaturation at the surface. The
parameters $A$ and $\sigma _{0}$ are different for the basal and prism
facets, and will generally depend on temperature, but they do not depend on $%
\sigma _{surf}$. We have measured $A(T)$ and $\sigma _{0}(T)$ over a fairly
broad range of temperatures for the two facets \cite{kglalphas13}, but we
often vary their values when investigating ice growth rates from other
experiments. We typically assume $\alpha =1$ for all nonfaceted surface
boundary pixels.

On the outermost facet surfaces (the top basal and prism terraces), we
reduce the supersaturation by the Gibbs-Thomson effect to%
\[
\sigma _{surf}\rightarrow \sigma _{surf}-\frac{\delta }{R} 
\]%
where $R$ is an effective curvature term defined in \cite{kglca13}. Our code
makes a rather crude estimate of $R,$ only applied to the outermost faceted
surfaces, since we have not yet found a satisfactory algorithm for
estimating the surface curvature of complex structures in cellular automata.
We believe that the Gibbs-Thomson effect is negligible in many typical
experimental circumstances, but we include it mainly to suppress the growth
of one-pixel-wide structures at low supersaturations, as described in \cite%
{kglca13}.

\subsection{Initial Relaxation}

After defining the seed crystal geometry and other parameters, our model
first calculates the supersaturation field $\sigma \left( r,z\right) $
around the crystal. We assume a simple outer boundary condition $\sigma
=\sigma _{\infty }$, where $\sigma _{\infty }$ is a constant input
parameter. For pixels not on the crystal boundary, we iterate the relaxation
equation 
\[
\sigma (\tau +\Delta \tau )=\Delta \tau \left[ f_{-}(r)\sigma (r-\Delta
r)+f_{+}(r)\sigma (r+\Delta r)+\sigma (z-\Delta z)+\sigma (z+\Delta z)\right]
\]%
to solve the Laplacian, where $\Delta \tau =1/4$ and%
\[
f_{\pm }(r)=\left( 1\pm \frac{\Delta r}{2r}\right) 
\]%
as described in \cite{kglca13}. This step is applied by shifting the $\sigma
(r,z)$ matrix by one pixel in $\pm r$ and $\pm z$, and then adding the
results, handling everything in full matrix form for better computational
efficiency. Special handling of the $r=0$ and $z=0$ lines is described in 
\cite{kglca13}.

For boundary pixels we use (here for an $r$-type boundary pixel)%
\begin{equation}
\sigma (r_{bound},z,\tau +\Delta \tau )=\Delta \tau \left[ f_{+}\sigma
(r+\Delta r)+f_{-}\sigma _{solid}+\sigma (z+\Delta z)+\sigma (z-\Delta z)%
\right]
\end{equation}%
with \cite{kglca13} 
\begin{eqnarray}
\sigma _{solid} &=&\sigma (r_{bound})\left( 1-g\alpha \Delta \xi \right) 
\textrm{ } \\
\Delta \xi &=&\frac{c_{ice}v_{kin}}{c_{sat}}\frac{\Delta r}{D}=\frac{\Delta r%
}{X_{0}} \\
X_{0} &=&\frac{c_{sat}}{c_{ice}}\frac{D}{v_{kin}}=\sqrt{\frac{2\pi m}{kT}}D
\\
&\approx &\textrm{0.145 }\mu \textrm{m}\cdot \left( \frac{D}{D_{air}}\right) 
\sqrt{\frac{kT_{-15C}}{kT}}
\end{eqnarray}%
where $D_{air}\approx 2\times 10^{-5}$ m$^{2}/$sec is the diffusion constant
in air at a pressure of one bar. A corner boundary pixel (kink site), with
neighboring ice pixels in both $r$ and $z,$ will propagate using%
\begin{equation}
\sigma (r_{bound},z_{bound},\tau +\Delta \tau )=\Delta \tau \left[
f_{+}\sigma (r+\Delta r)+gf_{-}\sigma _{solid}+\sigma (z+\Delta z)+g\sigma
_{solid}\right]
\end{equation}%
We incorporated an additional weight factor $g=1/\sqrt{N}$ in this
expression that was not present in \cite{kglca13}, where $N$ is the number
of neighboring ice pixels (for example, $N=1$ for a facet boundary pixel,
and $N=2$ for a kink site). This factor allows us to define $\alpha =1$ at
kink sites, instead of $1/\sqrt{2}$ as was done in \cite{kglca13}. This
formalism does not handle the growth of pixels with $N>2$ with great
accuracy, but few such pixels are present in our simple ice growth models.

Note that in all relaxation calculations, we recompute $\alpha $ on the
boundary pixels for each relaxation iteration, so that $\alpha \left( \sigma
\right) $ relaxes together with $\sigma (r,z),$ thus allowing any desired
parameterization of $\alpha \left( \sigma \right) ,$ in contrast to \cite%
{kelly13}.

The initial relaxation of the $\sigma $ field is done while allowing no
growth of the seed crystal. The number of steps needed to relax to the
desired solution of the Laplacian scales as the square of the crystal size,
so we iterate the above equations $N_{steps}$ times, using%
\[
N_{steps}=(N_{speed}/200)\left\{ \left[ (ir_{\max }+iz_{\max })/2\right]
^{2}+N_{0}\right\} 
\]%
where $ir_{\max }$ and $iz_{\max }$ are the maximum indices for the ice
pixels, and $N_{0}=2000$. We have found that setting the parameter $%
N_{speed}=200$ gives a solution for $\sigma \left( r,z\right) $ that is
accurate to roughly one percent. Smaller values give less accurate results,
but with increased computational speed.

\begin{figure}[htb] 
  \centering
  \includegraphics[width=5.0in,keepaspectratio]{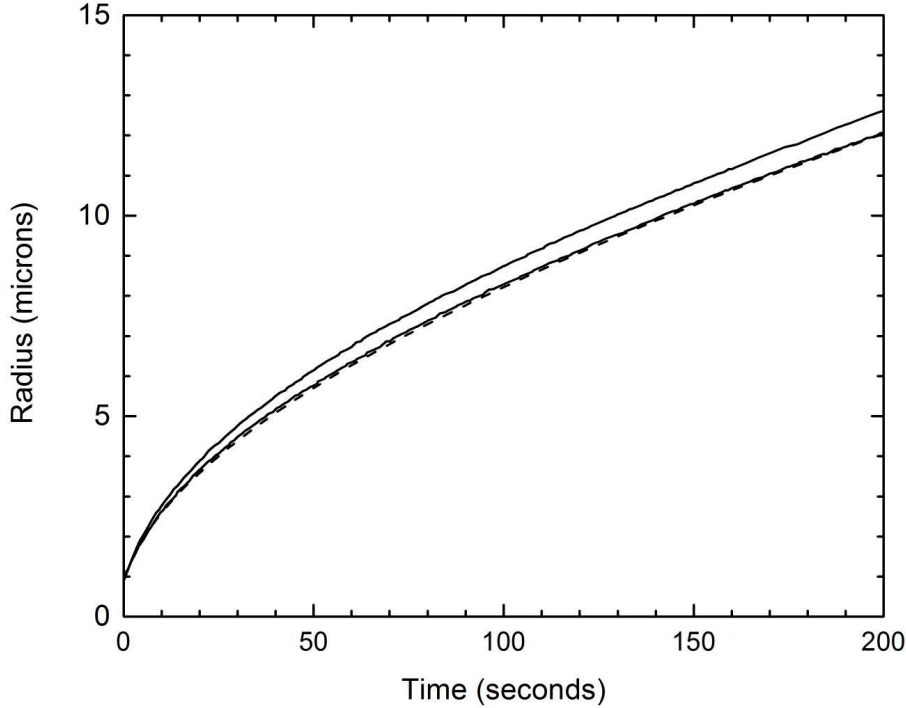}
  \caption{The dotted line shows the
growth of a spherical crystal, calculated analytically as described in 
\protect\cite{kglca13}. The two solid lines show the growth calculated with
our numerical model using $N_{speed}=50$ (upper solid line) and $%
N_{speed}=200$ (lower solid line). The 2D model in $(r,z)$ was constrained
to grow as a sphere for this test, as described in the text. This example
demonstrates the excellent quantitative agreement between theory and our
cellular automata model.}
  \label{sphere}
\end{figure}

\subsection{Crystal Growth}

We include crystal growth by defining an accumulated mass parameter $M$ for
each boundary pixel. We set $M=0$ when a pixel turns from an air pixel to a
boundary pixel (as the crystal grows), and a boundary pixel turns into an
ice pixel when $M$ reaches unity. Applying conservation of mass at the
boundary yields \cite{kglca13}%
\[
\frac{dM}{dt}=\frac{c_{sat}}{c_{ice}}\frac{Dg}{\Delta \xi X_{0}^{2}}\sum
\alpha \sigma f 
\]%
where the sum is over all neighboring ice pixels that drain the boundary
pixel. Here $\alpha $ and $\sigma $ are evaluated at the boundary pixel, $%
f=f_{\pm }$ for $r$ neighbors and $f=1$ for $z$ neighbors.

Following \cite{kelly13}, we separate the relaxation of the $\sigma $ field
from crystal growth. After relaxing the $\sigma $ field with the appropriate
boundary conditions, we then calculate $dM/dt$ for all the boundary pixels.
From this we advance the physical time by an amount $\Delta t$ such that $%
M\rightarrow 1$ for one, and only one, boundary pixel. This pixel is then
converted to ice, the other boundary pixels have their respective $M$
increased by the appropriate amounts, and the $\sigma $ field is again
relaxed before the next growth cycle. The relaxation step is essentially the
same as the initial relaxation described above, but using $N_{speed}$
iterations. This is much smaller than the number of iterations used in the
initial relaxation, because the perturbation of the $\sigma $ field is
rather small when a single boundary pixel turns to ice.

Since our goal was to produce a numerical growth model for quantitative
comparison with experimental data, we tested the model extensively using
analytic results for the diffusion-limited growth of simple morphologies, as
described in \cite{kglca13}. Figure \ref{sphere}, for example, shows our
model reproduction of the growth of a spherical crystal. Because a spherical
morphology is unstable to the Mullins-Sekerka growth instability, we
maintained the spherical shape by slightly adjusting the accumulated mass
parameter $M$ on the crystal boundary at each growth step, transferring mass
between boundary pixels while conserving total mass in the process. Figure %
\ref{sphere} demonstrates that our cellular automata model yields excellent
quantitative agreement with the analytic solution. Note that reducing $%
N_{speed}$ results in an incomplete relaxation of the supersaturation field,
and a corresponding increase in the crystal growth velocity.

\begin{figure}[htb] 
  \centering
  \includegraphics[width=5.67in,height=3.14in,keepaspectratio]{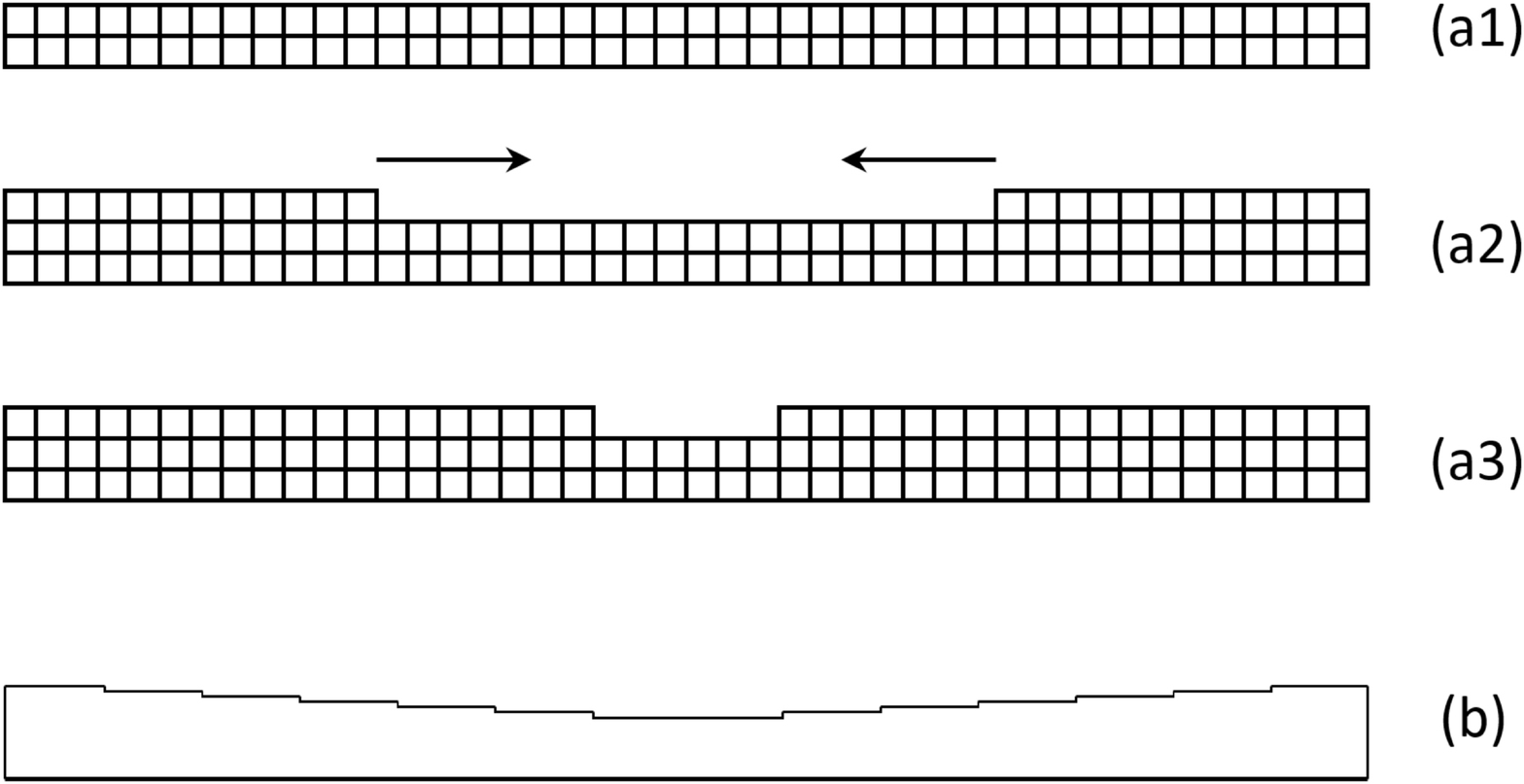}
  \caption{These sketches depict the
growth of a basal facet atop a columnar ice crystal, in all cases showing a
cross section of the facet in $(r,z)$ space, with $r=0$ at the center of the
sketches. The top three sketches (a1-a3) show our model crystal. It begins
as a perfectly faceted surface (a1), then a new terrace nucleates at the
edge of the facet (a2), and the new terrace grows inward (a2-a3). The bottom
sketch (b) shows our molecular picture of this same surface. New molecular
terraces again nucleate at the facet edge and grow inward, but many
molecular steps are present on the surface at all times. The coarse $z$
resolution of the cellular automata model limits our ability to model this
slightly concave faceted surface.}
  \label{facet}
\end{figure}

\subsection{Model Limitations}

It is instructive at this point to consider the differences between our
molecular picture of a growing faceted surface and the corresponding
macroscopic cellular automata model, as shown in Figure \ref{facet}. This
depicts, for example, a basal facet atop a columnar ice crystal. When there
are no available steps in the macroscopic model (a1), the growth is
determined by the nucleation of new layers set by $\alpha _{facet}.$ But
when a step appears, $\alpha $ increases substantially near the kink site.
If $x_{s}$ is larger than $(La/\Delta z)$, where $L$ is the size of the
model crystal, then $\alpha \rightarrow 1$ over the entire facet except on
the top terrace. In this new state (a2), the step grows quite rapidly across
the surface (a3), until it reaches the center of the facet and there are
again no steps.

In the molecular picture shown by sketch (b) in Figure \ref{facet}, the
surface includes a large number of molecular steps, and these steps all grow
simultaneously inward. The attachment coefficient is $\alpha \approx 1$ in a
strip roughly $x_{s}$ wide next to each molecular step, and is $\alpha
=\alpha _{facet}\ll 1$ otherwise. The overall growth velocity of the facet
is set by the step separation $x_{step}$ together with the nucleation of new
steps determined by $\alpha _{facet}.$ (Even this picture is too simple, as
step bunching \cite{saito96} will result in macrosteps on the ice surface,
which are often seen in ice growth experiments.)

An important difference in these two pictures is that the real crystal
contains multiple steps growing simultaneously, while the macroscopic model
exhibits slow faceted growth (with a slow increase in the accumulated mass
parameter) punctuated by the rapid completion of new terraces. In both
cases, the overall growth velocity is determined by the generation of new
terraces at the edge of the facet, and our model includes this key feature
of ice growth. However the model provides only an imperfect representation
of the growing faceted surface, as the model resolution $\Delta z$ is simply
not sufficient to represent the large number of simultaneous terraces
present on vicinal surfaces. Thus this model should work adequately in the
limit of layer-by-layer growth, and again for surfaces that are sufficiently
curved to include more than one $\Delta z$ step at all times. But vicinal
angles in the range $a/L<\theta <\Delta z/L$ cannot be modeled with perfect
fidelity. This shows that our inclusion of surface diffusion in the cellular
automata formalism has improved our ability to model faceted growth,
relative to a model with no surface diffusion. However, the coarse spatial
resolution means that this model still cannot be expected to reproduce real
crystal growth with perfect fidelity.

\subsection{Forced Faceting}

We found it useful to include an optional \textquotedblleft forced
faceting\textquotedblright\ feature in our model to partially address its
deficiencies when dealing with nearly faceted surfaces. When this feature is
turned on, we collect all the accumulated mass $dM$ on the top terrace at
each growth step and transfer it to the nearest inner kink site. This action
guarantees layer-by-layer growth while conserving mass in the process. The
top terrace $dM$ values are not transferred when the top terrace extends to $%
r=0$ (for basal facets) or $z=0$ (for prism facets). This action assumes
that new terraces nucleate at the outer edges of facets, which is true for
our simple growth morphologies.

We often use this feature when an experimental growth morphology is
perfectly faceted (to the resolution of our imaging) and the model produces
only a partially faceted morphology. The difference may lie in our
assumption of an infinite Ehrlich--Schwoebel barrier, since a leaky barrier
will promote faceting. But the difference may also come from our imperfect
modeling of nearly faceted surfaces, as described above. Our rationale for
including this model feature will become somewhat clearer when examining a
specific experimental case study.

\begin{figure}[htb] 
  \centering
  \includegraphics[width=5.67in,height=2.72in,keepaspectratio]{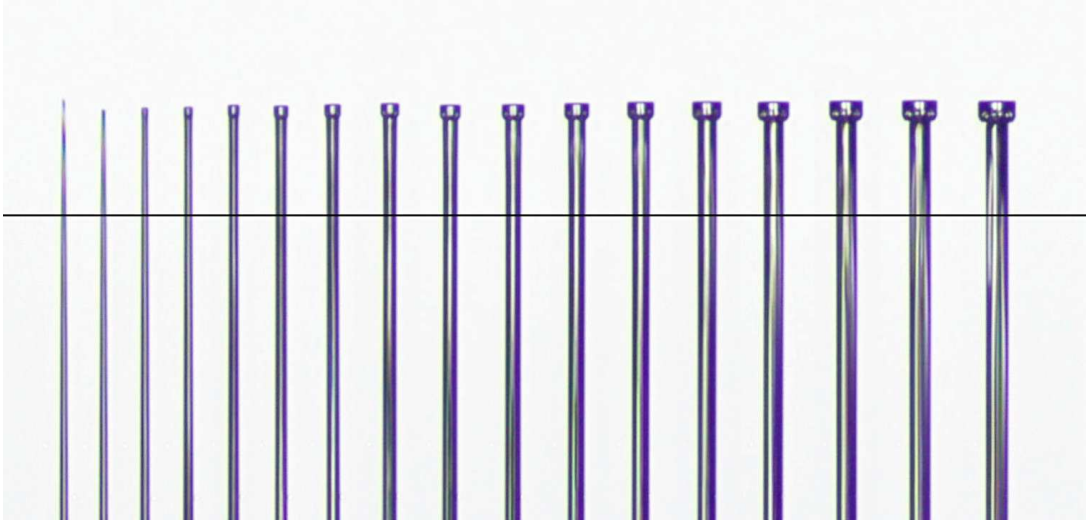}
  \caption{A composite image showing the
growth of a faceted block on the end of a thin ice needle as a function of
time. The ice crystal was grown in air at a temperature of -10 C, with a
supersaturation of $\protect\sigma _{\infty }\approx 11$ percent. The needle
axis is along the crystal's $c$ axis. The needle shrank slightly (first and
second images) as the temperature equilibrated in the growth chamber, and
grew thereafter. The total needle length was about two millimeters, and the
needle base provided a stable reference point for measuring the axial
growth. Scale for this image is provided by the measurements in Figure 
\protect\ref{data10}.}
  \label{tendata}
\end{figure}

\section{An Illustrative Example}

\begin{figure}[tb] 
  \centering
  \includegraphics[width=5.0in,keepaspectratio]{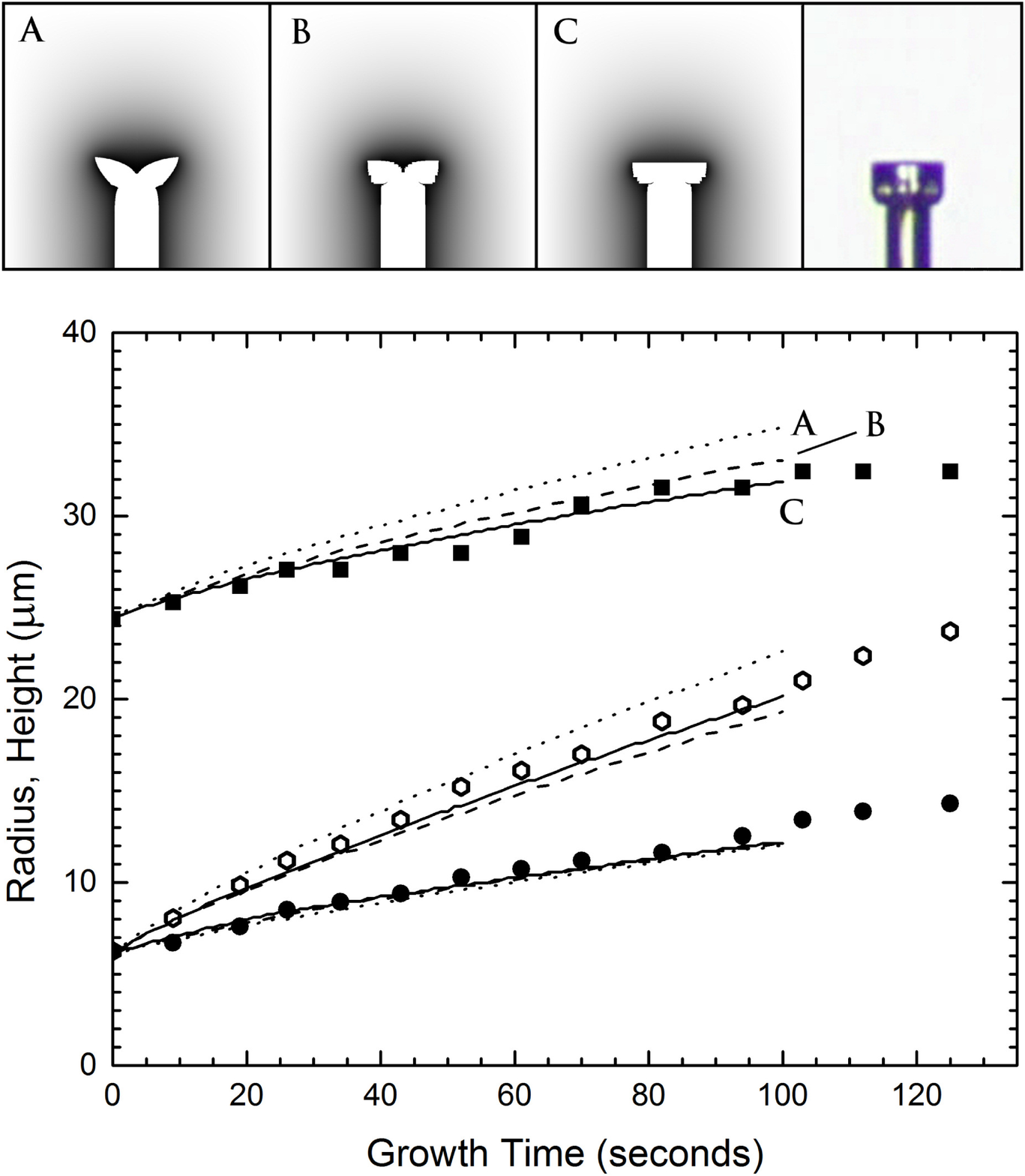}
  \caption{The data points show
measurements of the images in Figure \protect\ref{tendata}, giving the
needle radius $R_{needle}(t)$ (at a position 200 $\protect\mu $m below the
needle tip) (solid round points), the block radius $R_{block}(t)$ (open
round points) and the needle height $H(t)$ (solid square points) as a
function of time. The $H(t)$ points were shifted to give $H(0)=24$ $\protect%
\mu $m for plotting. The curves show models described in the text. The small
images show corresponding model cross sections at $t=100$ seconds, along
with the observed crystal at the same time and scale.}
  \label{data10}
\end{figure}

As our primary goal is the quantitative analysis of ice growth data, it is
instructive to examine some sample data in detail. Figure \ref{tendata}
shows a series of images of an ice needle as a function of time as it grew
in the apparatus described in \cite{kgldual14}. Analysis of these images
yielded the needle radius $R_{needle}(t)$, block radius $R_{block}(t)$, and
needle height $H(t)$ as a function of time, as shown in Figure \ref{data10}.
The radii were defined to give the same area as the hexagonal cross section
of the corresponding needle or block, and the time axis in the plot was
shifted to yield a (slightly extrapolated) straight needle at $t=0.$ The
total needle length was measured relative to a fixed reference at the needle
base (not shown in Figure \ref{tendata}), and $H(t)$ was shifted to give an
arbitrary $H=24$ $\mu $m at $t=0.$

To model these data, we assume that the attachment coefficients on the facet
surfaces have the form $\alpha _{facet}=A\exp (-\sigma _{0}/\sigma ),$ as
described in the previous section, and for nonfaceted sites we assume $%
\alpha =1.$After examining numerous models using a range of input
parameters, we found that the needle radius $R_{needle}(t)$ was an
especially good indicator of the supersaturation $\sigma _{\infty }$ far
from the growing crystal. We found that $R_{needle}(t)$ was roughly
proportional to $\sigma _{\infty }$, plus $R_{needle}(t)$ was rather
insensitive to $\alpha _{basal}$ and $\alpha _{prism}.$ This behavior arises
because the needle growth is largely diffusion limited, as can be seen from
the analytic solution for the growth velocity $v=dR_{needle}/dt$ of an
infinitely long needle \cite{kglca13}, which gives%
\[
v=\frac{\alpha \alpha _{diffcyl}}{\alpha +\alpha _{diffcyl}}v_{kin}\sigma
_{Rout} 
\]%
where $\alpha =\alpha _{prism}$ is the attachment coefficient at the needle
surface (which is a prism facet), $\sigma _{Rout}$ is the supersaturation at
the outer boundary, and%
\[
\alpha _{diffcyl}=\frac{1}{B}\frac{X_{0}}{R_{needle}} 
\]%
with $B=\log (R_{out}/R_{needle})$. With $R_{needle}=10$ $\mu $m and
assuming $B\approx 8$, this gives $\alpha _{diffcyl}\approx 0.002,$ which is
quite small. This means that $\alpha _{diffcyl}\ll \alpha _{prism}$ is a
reasonable approximation, giving $v\approx \alpha _{diffcyl}v_{kin}\sigma
_{Rout}$ independent of $\alpha _{prism}.$

By running a series of models, we found that $\sigma _{\infty }\approx 2.5$
percent gave a good fit to $R_{needle}(t),$ a value that is considerably
less than the experimental value of $\sigma _{\infty }\approx 11$ percent.
This difference arises mainly because we typically use $R_{out}=72$ $\mu $m
in our models, giving $B\approx 2.5.$ The experimental outer boundary is not
precisely defined, but estimating $R_{out}=2$ cm gives $B\approx 8,$ and the
ratio of these explains most of the difference in $\sigma _{\infty }.$

Given the various experimental and model uncertainties, we adjusted $\sigma
_{\infty }$ at the outer boundary of our model space to best fit $%
R_{needle}(t),$ and we used $\sigma _{\infty }=2.5$ percent for all the
models shown in Figure \ref{data10}. Other model parameters that did not
change from model to model are $Z_{out}=116$ $\mu $m, $T=-10$ C, $%
R_{needle}(t=0)=6.2$ $\mu $m, $H(t=0)=50$ $\mu $m, $\delta =0.3$ nm, $%
n_{SD}=10^{4},$ and $N_{speed}=50.$

Model A shown in Figure \ref{data10} used the parameters $M_{A}=\{\alpha
_{basal}=[1,1],\alpha _{prism}=[1,0.7]\}$ and $A_{SD}=0$, where we have used
the shorthand notation $\alpha =[A,100\sigma _{0}].$ Choosing $A_{SD}=0$ for
both facets means that the model is essentially that described in \cite%
{kglca13}, with no surface diffusion terms. Model B was the same as Model A,
except that it used $A_{SD}=1$ on both facets. Model C was the same as Model
B, expect that it also used forced faceting on the basal surface.

From this example we see that adding surface diffusion promotes faceting.
Model A, with no surface diffusion, showed essentially no faceting, while
Model B was closer to having faceted basal and prism surfaces. Still, Model
B showed basal hollowing that was not seen in the experimental crystal. The
growth velocities were changed only a small amount, owing to the fact that
the growth was substantially diffusion limited.

\section{Discussion}

One clear conclusion from this investigation is that surface diffusion is
likely an important factor in modeling ice crystal growth using cellular
automata. Even if the diffusion length $x_{s}$ is smaller than the model
resolution $\Delta z,$ the effects of surface diffusion can extend to rather
large distances in the model, of order $n_{SD}=(x_{s}/a)$ pixels. This
complicates cellular automata models, as surface diffusion is a nonlocal
phenomenon. While the importance of surface diffusion follows from the
fairly simple argument described above, it was not included in earlier
cellular automata models of ice crystal growth.

We implemented what we believe is a fairly accurate approximation of surface
diffusion by simply increasing $\alpha $ at surface points that are near
kink sites, as prescribed above, ignoring mass transfer to the kink sites.
This works because increasing the accumulated mass near kink sites has
essentially the same effect on the overall growth behavior as transfering
the accumulated mass to the relevant kink sites. Calcuating the necessary $%
\alpha _{SD}$ depends only on the crystal geometry at a given time, and does
not require much additional computation time.

When ice crystals are grown in air near one bar, which is a common
experimental condition, faceted surfaces are generally not faceted at the
molecular level, but are slightly concave. New layers nucleate at the facet
edges, where the supersaturation is highest, and terraces grow inward from
the edges. In many realistic circumstances, this means that $\alpha \approx
1 $ at essentially all points on the crystal surface, except for the top
terraces, where $\alpha =\alpha _{facet}.$ Moreover the top terraces might
be exceedingly narrow for typical facets -- a facet with a 1 $\mu $m
depression over a 100 $\mu $m width means that the top terraces are only 100
molecules wide. This makes for a somewhat remarkable circumstance --
apparently quite common in ice growth from vapor -- where $\alpha \approx 1$
everywhere on a complex faceted ice crystal except for a few extremely
narrow terraces where $\alpha \approx \alpha _{facet}\ll 1.$

Another conclusion from this investigation is that accurately determining
the intrinsic $\alpha _{facet}(\sigma ,T)$ for faceted ice surfaces is
extremely challenging when using only growth measurements made in air. The
effects of bulk diffusion together with surface diffusion are subtle and
difficult to model accurately, making it impractical in many circumstances
to extract $\alpha _{facet}(\sigma ,T)$ with any real accuracy. Fortunately,
observing growth in near vacuum seems to alleviate these problems
sufficiently to allow accurate measurements of $\alpha _{facet}(\sigma ,T)$ 
\cite{kglalphas13}. Assuming that the $\alpha _{facet}(\sigma ,T)$ values
determined at pressures near $0.01$ bar apply to pressures near 1 bar, these 
$\alpha _{facet}(\sigma ,T)$ can then be used as input to the numerical
models to further investigate structure formation that arises during
diffusion-limited growth at higher pressures.

\bibliography{C:/Dropbox/1-kgl-top/Papers/1Bibliography/kglbiblio3}
\bibliographystyle{phreport} 
\end{document}